# New equivalent model of quantizer with noisy input and its application for ADC resolution determination in an uplink MIMO receiver

Arkady Molev-Shteiman, Xiao-Feng Qi, Laurence Mailaender, Narayan Prasad

Radio Algorithms Research, Futurewei Technologies, Bridgewater, NJ, USA

Bertrand Hochwald,

Dept. of Electrical Engineering, University of Notre Dame, Notre Dame, IN, USA

*Abstract* – **When a quantizer input signal is the sum of the desired signal and input white noise, the quantization error is a function of total input signal. Our new equivalent model splits the quantization error into two components: a non-linear distortion (NLD) that is a function of only the desired part of input signal (without noise), and an equivalent output white noise. This separation is important because these two terms affect MIMO system performance differently. This paper introduces our model, and applies it to determine the minimal Analog-to-Digital Converter (ADC) resolution necessary to operate a conventional MIMO receiver with negligible performance degradation. We also provide numerical simulations to confirm the theory. Broad ramifications of our model are further demonstrated in two companion papers presenting low-complexity suppression of the NLD arising from insufficient ADC resolution, and a digital dithering that significantly reduces the MIMO transmitter Digital-to-Analog Converters (DAC) resolution requirement.**

# I. INTRODUCTION

Massive MIMO is an emerging technology capable of improving spectral efficiency of wireless communication by orders of magnitude [1]. However, a significant increase in base station antennas implies a proportional increase in cost and power consumption. On the other hand, it was shown that massive MIMO may significantly mitigate the impact of imperfections in the hardware implementation [2], implying that we may use cheaper and lower energy components to implement Massive MIMO. In the overall cost and energy budget of massive MIMO base station quantizers (ADCs and DACs) are important elements. It is known that power consumption of the quantizer grows exponentially with the number of quantization bits [3]. Therefore algorithms that reduce quantizer resolution have significant practical importance.

Intuitively, the quantizer resolution should be sufficiently high to ensure that quantization noise power is sufficiently below that of thermal noise. As the received SNR at each MIMO observation (antenna) decreases with an increase of observation (antenna) number, the thermal noise power at the ADC input grows. It follows that the tolerable quantization error increases and quantizer resolution decreases accordingly, to a single bit in the extreme.

The conventional approach assume that ADC resolution is sufficiently high so we may approximate quantization error as independent white process with variance $\Delta^2/12$ equal where $\Delta$ is the quantization step [4]. In such cases we can set the ADC resolution to ensure that quantization noise variance is much lower that thermal noise variance. However for low resolution ADCs this approximation is not accurate and therefore for Massive MIMO, ADC resolution determination is much more challenging. Also in Massive MIMO, optimal setting of ADC resolution is much more critical and excessive safety margin is not an option,

The problem of low-resolution AD/DA in context of multi antenna Massive MIMO has raised a lot of discussion. Many contributions which consider uplink Massive MIMO receivers with arrays of low-resolution ADCs have been published. A general overview of low resolution ADC is given in [8]. An information theoretic analysis of Massive MIMO uplink receiver with low resolution ADC performance is given in [9]-[27]. Other works propose different methods of data reception [28]-[35] and channel estimation [36]-[41] taking into account limited ADC resolution. The problem of low resolution DAC has also attracted research attention recently. Different methods of low resolution DAC precoding were presented in [43]-[49].

We identify three main methods of dealing with finite resolution ADCs as follows:

- Additive Quantization Noise Model (AQNM) [4] that represents the ADC output as the sum of ADC input and quantization error which is uncorrelated with ADC output. This model is correct only if the expectation of the ADC input given ADC output is equal to ADC output, which implies a specially-designed ADC, designed to match input Probability Density Function (PDF), like the Lloyd-Max quantizer [4]. For the commonly used uniform ADC, this approximation is only valid at high resolution (see [9]-[14] and [28]-[31]).

- Probabilistic methods that search for the desired input signal which maximizes the likelihood of the observed ADC output vector (see [15]-[24], [32]-[35] and [36]-[40]).

- Bussgang-Rowe decomposition method [5]-[6] that considers the ADC as a non-linear element. It presents the ADC output as the sum of a scaled version of the input signal and a NLD that is uncorrelated with input signal (see [25]-[27] and [41]).

Our method also uses the Bussgang decomposition, however instead of applying Bussgang to the original quantizer transfer function, we apply it to an equivalent transfer function which is given by the expectation over the noise of the ADC output given the input desired signal. We also replace the additive noise at the ADC input by its equivalent on the ADC output. Unlike the conventional Bussgang decomposition for which NLD is a function of the total input signal, the equivalent decomposition decouples the noise from the desired signal, and presents NLD as a function of only the desired part of input signal (without noise).

This new approach provides the following benefits:

- It decomposes the quantization error into two components: the equivalent white noise whose autocorrelation is the Dirac delta function, and the non-white NLD whose autocorrelation may differ from the Dirac delta function. This separation is important because these two components affect the MIMO system differently. Using this new model, we show how to determine the minimum ADC resolution required for a conventional massive MIMO receiver to operate with performance degradation below a desired value. To the best of our knowledge we are the first to determine this sufficient resolution.

- According to our model, the NLD is a function of desired signal only. Therefore when the ADC resolution is insufficient we may design low complexity algorithms for NLD suppression. The companion paper [55] presents an example of such an algorithm.

- Since we know the dependency of this equivalent transfer function on input noise statistics, we may design an optimal dither to mitigate the NLD effect. The companion paper [56] presents an example of such a digital dither for a MIMO downlink transmitter DAC resolution reduction.

- In contrast to existing approaches, our formulation allows a visual, intuitive understanding of the dependence of the equivalent ADC/DAC transfer function on input noise statistics, which allow us to actually design and implement practical low-resolution transceivers.

The validity of our theory has been confirmed by the fact that many conclusions that were reached by other methods also follow from our equivalent model. For instance, the well-known fact that (see, e.g. [20], [21] and [23]) in low SNR regime we may use a conventional MIMO receiver without taking into account ADC NLD, however in the high SNR regime this NLD has to be explicitly taken into account. The model also arrives at the well-known fact (see, e.g. [23] and [41]) that in the low SNR regime, a one-bit ADC causes $\pi/2 = 1.96$ dB performance degradation relative to its infinite-resolution counterpart. Likewise the channel estimation algorithms that were derived by a probabilistic method (see, e.g. [36] and [37]) also follow from our model.

The fact that additive noise (dither) mitigates the effect of NLD has been known for over fifty years. The effect of dither on the equivalent transfer function of a non-linear element was shown in [51]-[53]. However we believe that we are the first to apply the equivalent transfer function to the Bussgang decomposition and bring it to bear on the MIMO problem.

A MIMO receiver estimates the vector of transmitted user's signals (Multiple Inputs) based on vector of received signal observations (Multiple Output). However, it appears that the MIMO principle broadly applies to communication schemes beyond those operating in the spatial domain. For example, a Code Division Multiple Access (CDMA) receiver also estimates vector of transmitted user's signals based on vector of received signal observations, though these observations are obtained in time domain. Our method of MIMO quantizer analysis is valid for MIMO models in any domain. However our primary target is multi antenna MIMO, where the problem of quantizer resolution reduction is most urgent.

## II. SYSTEM MODEL

An ADC performs the following quantization operation:

$$Q(s) = \begin{cases} +(R-1)\cdot \Delta/2 & IF\ (s \geq +(R-2)\cdot \Delta/2) \\ -(R-1)\cdot \Delta/2 & IF\ (s < -(R+2)\cdot \Delta/2) \\ \Delta \cdot round((s/\Delta)+0.5)-1 & ELSE \end{cases} \quad (0.1)$$

where $\Delta$ is the quantization step size, $round(\ )$ denotes rounding operation, and $R$ is the total number of possible quantizer output level $q_r = (2\cdot r - R - 1)\cdot \Delta/2$ for $(r = 1, 2, ..., R)$. The quantizers bits number is equal to $\log_2(R)$. We assume without loss of generality that $\Delta = 2$. As an example, the 2-bit quantizer transfer function is shown in Figure 1:

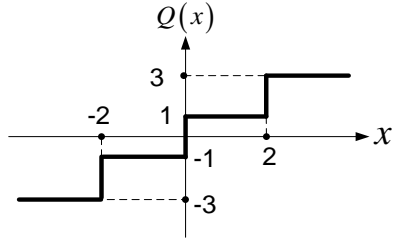

Figure 1: 2-bit quantizer transfer function

We define as uplink MIMO communication system over a flat-fading channel as follows.

$$\tilde{S}_O = \tilde{Q}(\tilde{S}_I + \tilde{N}_I) \quad \text{where} \quad \tilde{S}_I = \tilde{H} \cdot \tilde{X}_I \quad (0.2)$$

where we use a superscript '~' to denote a complex quantity, $\tilde{S}_I$ is the desired signal vector that we wish to pass through ADCs with minimum distortion, $\tilde{N}_I$ is the input noise vector, $\tilde{S}_O$ is the output observations (received signal) vector, $\tilde{X}_I$ is the input (user signal) vector, $H$ is the channel matrix and $\tilde{Q}(\ )$ denotes element-wise complex quantization operation:

$$\tilde{Q}(\tilde{s}) = Q(\text{Re}(\tilde{s})) + j \cdot Q(\text{Im}(\tilde{s})) \quad \text{where} \quad j = \sqrt{-1} \tag{0.3}$$

Vectors $\tilde{S}_I, \tilde{S}_O, \tilde{N}_I$ have length $M$, where $M$ is the number of observations, vector $\tilde{X}_I$ has length $K$, where $K$ is the number of users, and channel matrix $H$ has size $M \times K$.

We denote an element of vectors $\tilde{X}_I, \tilde{S}_I, \tilde{S}_O, \tilde{N}_I$ and matrix $H$ as $\tilde{x}_k, \tilde{s}_I(m), \tilde{n}_I(m), \tilde{s}_O(m)$ and $\tilde{h}_k(m)$, where $k$ is the user index $(k=1,...,K)$ and $m$ is the observation index $(m=1,...,M)$.

For multi antenna MIMO systems, the observations vector is the vector of antenna domain samples and $M$ is the total number of antennas. For CDMA system, the observations vector is the vector of time domain samples and $M$ is the total number of chips of the spreading code.

We assume for simplicity that there is no multipath and the channel is line of sight (LOS):

$$\tilde{h}_k(m) = \tilde{g}_k \cdot \tilde{c}_k(m) \tag{0.4}$$

where $\tilde{g}_k$ is the user $k$ channel complex amplitude and $\tilde{c}_k(m)$ is the user $k$ antenna $m$ phase shift coefficient satisfying, $|\tilde{c}_k(m)| = 1$ where $|\ |$ denotes absolute value operation.

For CDMA $\tilde{c}_k(m)$ is the chip $m$ of user $k$'s spreading code. For multi antenna MIMO $\tilde{b}_k(m)$ is the steering coefficient of user $k$ at antenna $m$. The actual realization of the steering function depends on an antenna array configuration. For example, the steering function of MIMO receiver equipped with uniform linear antenna array is given by:

$$\tilde{c}_k(m) = \exp(j \cdot \pi \cdot m \cdot \sin(\alpha_k)) \quad \text{where} \quad \alpha_k \text{ is the angle of arrival of user } k \tag{0.5}$$

According to (0.4) we rewrite MIMO equations (0.2) in scalar form as,

$$\tilde{s}_O(m) = \tilde{Q}(\tilde{s}_I(m) + \tilde{n}_I(m)) \quad \text{where:} \quad \tilde{s}_I(m) = \sum_{k=1}^{K} \tilde{g}_k \cdot \tilde{c}_k(m) \cdot \tilde{x}_k \tag{0.6}$$

Let us allocate indices $2 \cdot m - 1$ and $2 \cdot m$ to the real and imaginary part of the complex observation $m$, respectively, and rewrite the system model in terms of real-valued quantities only:

$$s_I(2 \cdot m - 1) = \text{Re}(\tilde{s}_I(m)) \quad s_I(2 \cdot m) = \text{Im}(\tilde{s}_I(m))$$
$$n_I(2 \cdot m - 1) = \text{Re}(\tilde{n}_I(m)) \quad n_I(2 \cdot m) = \text{Im}(\tilde{n}_I(m)) \quad (0.7)$$
$$s_O(2 \cdot m - 1) = \text{Re}(\tilde{s}_O(m)) \quad s_O(2 \cdot m) = \text{Im}(\tilde{s}_O(m))$$

With this notation we may rewrite expression (0.6) for real observations as:

$$s_O(m) = Q(s_I(m) + n_I(m)) \quad (0.8)$$

We assume that the additive noise $n_I(m)$ on each observation is a real independent Gaussian random variable with zero mean and identical variance $\sigma_N^2$. We denote its normalized (unit variance) Probability Density Function (PDF) as $p_N(x) = \exp(-x^2/2)/\sqrt{2 \cdot \pi}$.

We assume that each user signal $\tilde{x}_k$ is circular-symmetric complex Gaussian random variable with variance $2 \cdot \sigma_X^2$. From this assumption and (0.6) it follows that each desired signal $\tilde{s}_I(m)$, being the superposition of many weighted circular-symmetric complex Gaussian random variables, is also a circular-symmetric complex Gaussian random variable.

According to (0.6), the real and imaginary parts of each $\tilde{s}_I(m)$ have the same variance,

$$\sigma_S^2 = E\left[s_I(m)^2\right] = \sum_{k=1}^{K} |\tilde{g}_k|^2 \cdot \sigma_X^2 \quad \text{where } E[\ ] \text{ denotes expectation.} \quad (0.9)$$

We denote normalized PDF of desired signal $s_I(m)$ as: $p_S(x) = \exp(-x^2/2)/\sqrt{2 \cdot \pi}$.

We define the ADC input Signal to Noise Ratio (SNR) as the ratio between desired signal and noise variances and the ADC Scaling Factor (SF) as the ratio between ADC saturation level and input signal variance,

$$SNR_{ADC,In} = \sigma_S^2/\sigma_N^2 = \sum_{k=1}^{K} |\tilde{g}_k|^2 \cdot \sigma_X^2/\sigma_N^2 \qquad SF = R^2/(\sigma_S^2 + \sigma_N^2) \quad (0.10)$$

We define the cumulative input SNR as the sum of input SNRs of all ADCs in the array:

$$SNR_{\Sigma,In} = \sum_{m=1}^{M} SNR_{ADC,In} = M \cdot SNR_{ADC,In} \quad (0.11)$$

Any MIMO receiver detects a user signal vector $\tilde{X}_I$ with a certain estimation error. Let us define the ADC array performance degradation - Noise Figure (NF) as the ratio between the estimation error variances of MIMO receivers equipped with finite resolution ADCs and with ideal ADCs (no quantization error). Let us define the worst case NF as the maximum NF over all users and all possible channel realizations:

$$NF_{MAX} = \max_{H,k}\left( E\left[\left|\tilde{x}_k - \hat{\tilde{x}}_{k,Q}\right|^2\right] \Big/ E\left[\left|\tilde{x}_k - \hat{\tilde{x}}_{k,I}\right|^2\right] \right) \quad (0.12)$$

where $\hat{\tilde{x}}_{k,I}$ is the ideal receiver's user signal estimate and $\hat{\tilde{x}}_{k,Q}$ is the quantized receiver's user signal estimate.

As we will show in the sequel the worst case NF is a decreasing function of ADC resolution. The goal of our paper is to determine the minimal ADC resolution required to keep the NF below a certain limit.

The most typical receiver options for detecting user signal vector $\tilde{X}_I$ without concern for quantization error are shown in Table 1, where $(\ )^H$ denotes matrix conjugate transpose operation, $inv(\ )$ denotes matrix inversion operation, $diag(\ )$ denotes matrix diagonalization operation and *I* is the Identity Matrix. The table includes Maximum Ratio Combining (MRC), Maximum Likelihood (ML), Zero Forcing (ZF) and Minimum Mean Square Error (MMSE) receivers. As we may see from this table the MRC signal $\tilde{X}_{MRC}$ is the common front-end for the other receiver types. We may always model the MRC estimation error variance increase as an equivalent increase of input additive white noise. Therefore the NF of the MRC receiver upper bounds the NF of the other receiver types just mentioned. Consequently, if the ADC resolution is sufficient for the MRC receiver, it will also be sufficient for the other receiver types. For this reason, our paper only evaluates the NF of the MRC receiver.

| Receiver Type | Operation |
|---|---|
| MRC | $\tilde{X}_O = inv\left(diag\left(\tilde{H}^H \cdot \tilde{H}\right)\right) \cdot \tilde{X}_{MRC}$    where    $\tilde{X}_{MRC} = \tilde{H}^H \cdot \tilde{S}_O$ |
| ML | $\tilde{X}_O = \arg\min_{\tilde{X}}\left(\left(\tilde{X}_{MRC} - \tilde{H}^H \cdot H \cdot \tilde{X}\right)^H \cdot inv\left(\tilde{H}^H \cdot \tilde{H}\right) \cdot \left(\tilde{X}_{MRC} - \tilde{H}^H \cdot H \cdot \tilde{X}\right)\right)$ |
| ZF | $\tilde{X}_O = inv\left(\tilde{H}^H \cdot \tilde{H}\right) \cdot \tilde{X}_{MRC}$ |
| MMSE | $\tilde{X}_O = inv\left(\tilde{H}^H \cdot \tilde{H} + I \cdot \sigma_N^2 / \sigma_X^2\right) \cdot \tilde{X}_{MRC}$ |

Table 1    The different receiver types.

The MRC receiver without concern for quantization error estimates user signal $\tilde{x}_k$ as,

$$\hat{\tilde{x}}_k = MRC_k\left(\tilde{S}_O\right) = \left(1 \Big/ \sum_{m=1}^{M}\left|\tilde{h}_k(m)\right|^2\right) \cdot \sum_{m=1}^{M} \tilde{h}_k(m)^* \cdot \tilde{s}_O(m) = \frac{\tilde{g}_k^*}{M \cdot \left|\tilde{g}_k\right|^2} \cdot \sum_{m=1}^{M} \tilde{c}_k(m)^* \cdot \tilde{s}_O(m) \quad (0.13)$$

where $(\ )^*$ denotes the complex-conjugate operation.

Let us define the user *k* input SNR as the output SNR of the ideal MRC receiver,

$$SNR_{User,In}(k) = E\left[\left|\tilde{x}_k\right|^2\right] \Big/ E\left[\left|\tilde{x}_k - \hat{\tilde{x}}_k\right|^2\right] = M \cdot \left|\tilde{g}_k\right|^2 \cdot \sigma_X^2 / \sigma_N^2 \quad (0.14)$$

From the above it follows that cumulative input SNR is equal to sum of all users input SNRs,

$$SNR_{\Sigma,In} = M \cdot SNR_{ADC,In} = \sum_{k=1}^{K} SNR_{User,In}(k) \quad (0.15)$$

### III. Equivalent Model of Quantizer with Noisy Input

According to (0.8), the quantizer input is the sum of desired signal $s_I(m)$ and $n_I(m)$. Our goal is to pass $s_I(m)$ through the quantizer with minimum distortion. Let us define the expectation of noisy quantizer output $s_O$ given ADC input $s_I$ as the quantizer equivalent transfer function $F(\sigma_N, s_I)$. As we will see below, this particular definition of the transfer function results in an error signal that is white and uncorrelated with the quantizer input. It equals to the time-reversed convolution of real quantizer transfer function $Q(\ )$ and noise PDF $p_N(\ )$:

$$F(\sigma_N, s_I) = E[s_O | s_I] = \int_{x=-\infty}^{\infty} Q(s_I + \sigma_N \cdot x) \cdot p_N(x) \cdot dx \tag{0.16}$$

Let us define the quantizer equivalent noise as,

$$n_O(m) = s_O(m) - F(\sigma_N, s_I(m)) \tag{0.17}$$

From (0.16) and (0.17) it follows that:

$$E\left[n_O(m)^2 | s_I(m)\right] = E\left[s_O(m)^2 | s_I(m)\right] - F(\sigma_N, s_I(m))^2 = V(\sigma_N, s_I) - F(\sigma_N, s_I(m))^2 \tag{0.18}$$

where the energy function $V(\sigma_N, s_I)$ is defined as,

$$V(\sigma_N, s_I) = E\left[s_O^2 | s_I\right] = \int_{x=-\infty}^{\infty} Q(s_I + \sigma_N \cdot x)^2 \cdot p_N(x) \cdot dx \tag{0.19}$$

Therefore we may express the equivalent noise variance as:

$$\sigma_{NO}(\sigma_N, \sigma_S)^2 = E\left[n_O(m)^2\right] = \int_{x=-\infty}^{\infty} \left(V(\sigma_N, \sigma_S \cdot x) - F(\sigma_N, \sigma_S \cdot x)^2\right) \cdot p_S(x) \cdot dx \tag{0.20}$$

From the quantizer equivalent noise definition (0.17) it follows that,

$$E\left[n_O(m) | s_I(m)\right] = E\left[s_O(m) - F(\sigma_N, s_I(m)) | s_I(m)\right] = E\left[s_O(m) | s_I(m)\right] - E\left[s_O(m) | s_I(m)\right] = 0 \tag{0.21}$$

Because $n_O(m)$ is a function of $s_I(m)$ and $n_I(m)$ only and because $n_I(m)$ is independent random variable, and according to (0.21):

$$E\left[n_O(m) | s_I(m), s_I(n)\right] = E\left[n_O(m) | s_I(m)\right] = 0 \tag{0.22}$$

Therefore according to (0.21) for any pair $(m \neq n)$:

$$E\left[n_O(m) \cdot s_I(n) | s_I(m), s_I(n)\right] = E\left[n_O(m) | s_I(m), s_I(n)\right] \cdot s_I(n) = 0 \cdot s_I(n) = 0 \tag{0.23}$$

$$E\left[n_O(m) \cdot n_O(n) | s_I(m), s_I(n)\right] = E\left[n_O(m) | s_I(m)\right] \cdot E\left[n_O(n) | s_I(n)\right] = 0 \cdot 0 = 0 \tag{0.24}$$

If a conditional expectation always equals zero, then the unconditional expectation is also zero.

Then from (0.21), (0.23) and (0.24) it follows that the equivalent noise has zero mean, zero correlation with input signal and its autocorrelation is the Dirac delta function (white process):

$$E\left[n_O(m)\right] = 0 \quad \text{for any } m \tag{0.25}$$

$$E\left[n_O(m) \cdot s_I(n)\right] = 0 \quad \text{for any pair of } m \text{ and } n \tag{0.26}$$

$$E\left[n_O(m) \cdot n_O(n)\right] = 0 \quad \text{for any } m \neq n \tag{0.27}$$

We thus obtained the equivalent model of the quantizer with noisy input. We may represent the output of such a quantizer as the sum of the desired signal that passes through the equivalent transfer function and the equivalent additive white noise at the output.

$$s_O(m) = Q\left(s_I(m) + n_I(m)\right) = F\left(\sigma_N, s_I(m)\right) + n_O(m) \tag{0.28}$$

A block diagram of the noisy quantizer equivalent model is shown in Figure 2.

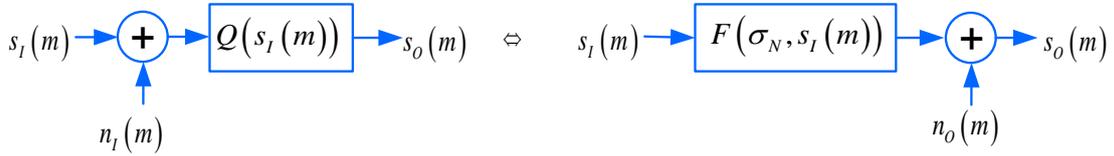

Figure 2  Equivalent block diagram of quantizer with noisy input

Assuming the desired signal $s_I(m)$ has a Gaussian distribution we may apply the Bussgang-Rowe decomposition [5],[6] to represent the output of a non-linear element as a sum of desired signal $s_I(m)$ with a certain gain $g_O(\sigma_N, \sigma_S)$ and an NLD $w_O(m)$.

$$F\left(\sigma_N, s_I(m)\right) = g_O(\sigma_N, \sigma_S) \cdot s_I(m) + w_O(m) \tag{0.29}$$

$$g_O(\sigma_N, \sigma_S) = E\left[s_I \cdot F(s_I)\right] / E\left[s_I^2\right] = \frac{1}{\sigma_S^2} \cdot \int_{x=-\infty}^{+\infty} F(\sigma_N, \sigma_S \cdot x) \cdot \sigma_S \cdot x \cdot p_S(x) \cdot dx \tag{0.30}$$

$$w_O(m) = F\left(\sigma_N, s_I(m)\right) - g_O(\sigma_N, \sigma_S) \cdot s_I(m) \tag{0.31}$$

The NLD term $w_O(m)$ has properties,

$$E\left[s_I(m) \cdot w_O(n)\right] = 0 \quad \text{for any pair } m \text{ and } n \tag{0.32}$$

$$E\left[w_O(m)\right] = 0 \quad \text{for any } m \tag{0.33}$$

Its variance is equal to,

$$\sigma_{WO}(\sigma_N, \sigma_S)^2 = E\left[w_O(m)^2\right] = \int_{x=-\infty}^{+\infty} F(\sigma_N, \sigma_S \cdot x)^2 \cdot p_S(x) \cdot dx - g_O(\sigma_N, \sigma_S)^2 \cdot \sigma_S^2 \tag{0.34}$$

According to (0.22) and (0.31),

$$E\left[n_o(m)\cdot w_o(n)|s_I(m),s_I(n)\right]=E\left[n_o(m)|s_I(m),s_I(n)\right]\cdot\left(F(\sigma_N,s_I(m))-g_o(\sigma_N,\sigma_S)\cdot s_I(m)\right)=0 \quad (0.35)$$

If a conditional expectation always equals zero, then the unconditional expectation is also zero.

$$E\left[n_o(n)\cdot w_o(m)\right]=0 \quad \text{for any pair } m \text{ and } n \quad (0.36)$$

However we cannot assume that the NLD is necessarily white, $E\left[w_o(m)\cdot w_o(n)\right]\neq 0$.

To summarize: The output of the quantizer may be represented as the sum of a desired signal multiplied by an equivalent gain, additive white noise and non-white NLD.

$$s_O(m)=g_o(\sigma_N,\sigma_S)\cdot s_I(m)+w_O(m)+n_o(m) \quad (0.37)$$

In contrast to the conventional Bussgang approach, here $w_O(m)$ is a function of only the desired part of input signal $s_I(m)$ (without input noise $n_I(m)$).

## IV. EQUIVALENT ADC TRANSFER FUNCTION FOR GAUSSIAN INPUT NOISE DISTRIBUTION

When the additive noise PDF is Gaussian, the ADC equivalent transfer function (0.16) and energy function (0.19) are given by:

$$F(\sigma_N,s_I)=\sum_{r=1}^{R}q_r\cdot\Pr(s_O=q_r|s_I) \quad \text{and} \quad V(\sigma_N,s_I)=\sum_{r=1}^{R}q_r^2\cdot\Pr(s_O=q_r|s_I) \quad (0.38)$$

Where $erf(\ )$ denotes the error function and:

$$\Pr(s_O=q_1|s_I)=\Pr((s_I+n_I)<(q_1+1))=0.5\cdot\left(1+erf\left((q_1+1-s_I)\big/\sqrt{2\cdot\sigma_N^2}\right)\right)$$

$$\Pr(s_O=q_R|s_I)=\Pr((s_I+n_I)\geq(q_R-1))=0.5\cdot\left(1-erf\left((q_R-1-s_I)\big/\sqrt{2\cdot\sigma_N^2}\right)\right)$$

For any other $r=2,...,R-1$

$$\Pr(s_O=q_r|s_I)=\Pr((q_r-1)\leq(s_I+n_I)<(q_r+1))=0.5\cdot\left(erf\left(\frac{q_r+1-s_I}{\sqrt{2\cdot\sigma_N^2}}\right)-erf\left(\frac{q_r-1-s_I}{\sqrt{2\cdot\sigma_N^2}}\right)\right)$$

(0.39)

For the special case of a 1-bit ADC it is equal to,

$$F(\sigma_N,s_I)=erf\left(s_I\big/\sqrt{2\cdot\sigma_N^2}\right) \quad \text{and} \quad V(\sigma_N,s_I)=1 \quad (0.40)$$

Figure 3 presents the ADC equivalent transfer function for 1- and 2-bit ADCs, respectively, at different noise variances. It shows that additive noise with sufficient variance has a linearizing effect on the ADC equivalent transfer function.

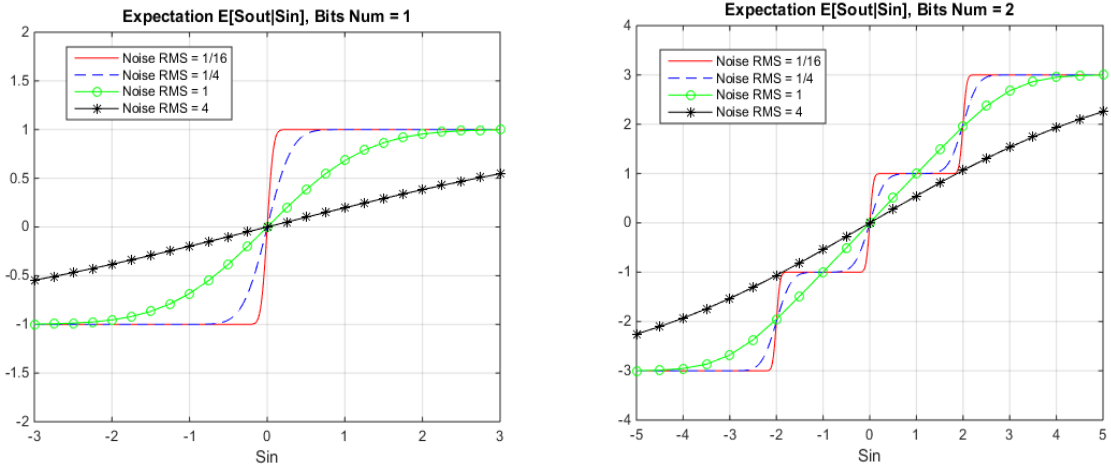

Figure 3    The equivalent transfer function of 1 bit (left) and 2 bit (right) ADCs

## V.  SINGLE ADC SNR, SDR AND SINAD FOR GAUSSIAN INPUT SIGNAL DISTRIBUTION

Let us define the ADC output SNR, SDR (Signal to Distortion Ratio) and SINAD (Signal to Noise and Distortion Ratio) as:

$$SNR_{ADC,Out}(\sigma_N, \sigma_S) = g_O(\sigma_N, \sigma_S)^2 \cdot \sigma_S^2 / \sigma_{NO}(\sigma_N, \sigma_S)^2 \tag{0.41}$$

$$SDR_{ADC,Out}(\sigma_N, \sigma_S) = g_O(\sigma_N, \sigma_S)^2 \cdot \sigma_S^2 / \sigma_{WO}(\sigma_N, \sigma_S)^2 \tag{0.42}$$

$$SINAD_{ADC,Out}(\sigma_N, \sigma_S) = g_O(\sigma_N, \sigma_S)^2 \cdot \sigma_S^2 / \left(\sigma_{NO}(\sigma_N, \sigma_S)^2 + \sigma_{WO}(\sigma_N, \sigma_S)^2\right) \tag{0.43}$$

According to (0.10) we may also express the ADC output SNR, SNR and SINAD as functions of ADC input SNR and SF, and here we show it both ways.

$$SNR_{ADC,Out}(SF, SNR_{ADC,In}) = SNR_{ADC,Out}\left(\sigma_N = R/\sqrt{SF \cdot (1+SNR_{ADC,In})}, \sigma_S = \sigma_N \cdot \sqrt{SNR_{ADC,In}}\right) \tag{0.44}$$

$$SDR_{ADC,Out}(SF, SNR_{ADC,In}) = SDR_{ADC,Out}\left(\sigma_N = R/\sqrt{SF \cdot (1+SNR_{ADC,In})}, \sigma_S = \sigma_N \cdot \sqrt{SNR_{ADC,In}}\right) \tag{0.45}$$

$$SINAD_{ADC,Out}(SF, SNR_{ADC,In}) = SINAD_{ADC,Out}\left(\sigma_N = R/\sqrt{SF \cdot (1+SNR_{ADC,In})}, \sigma_S = \sigma_N \cdot \sqrt{SNR_{ADC,In}}\right) \tag{0.46}$$

We used numerical integration to evaluate (0.44), (0.45) and (0.46) as functions of ADC input SNR and SF for a Gaussian input signal and noise distribution. The numerical integration interval was chosen from minus to plus $5 \cdot \sqrt{\sigma_S^2 + \sigma_N^2}$ with a step of $0.01 \cdot \min(\sigma_S, \sigma_N)$.

Figure 4 shows single ADC output SNR and SDR as functions of input SNR calculated for optimal SF that maximize ADC output SINAD(0.46):

$$SF_{Optim} = \arg\max_{SF} \left( SINAD_{ADC,Out} \left( SF, SNR_{ADC,In} \right) \right) \tag{0.47}$$

The optimal SF was obtained by searching over SF from 0 to 30dB with step 0.1dB.

From this figure we may conclude that the ADC output NLD power is a monotonically increasing function of input SNR until it converges to certain fixed value when the input SNR approaches infinity. In contrast, the ADC output white noise power is a monotonically decreasing function of input SNR. The consequences of this will be explained in section VII

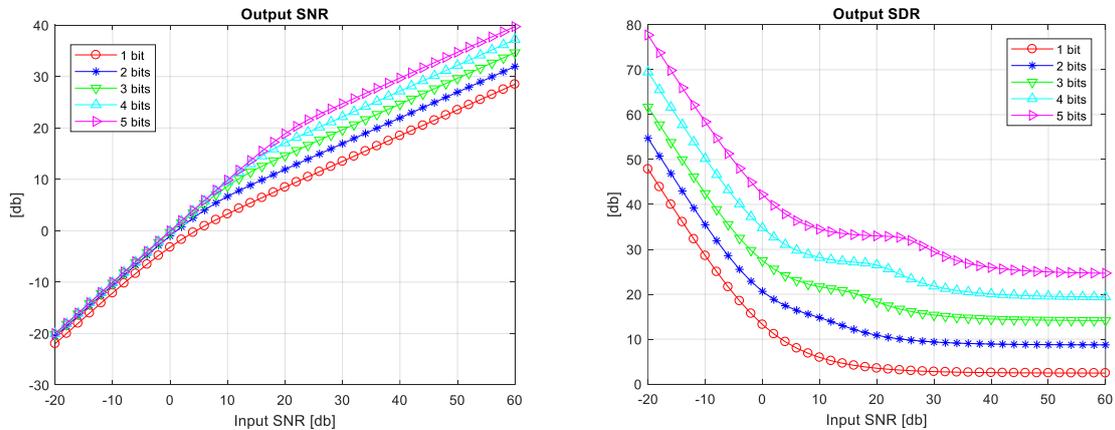

Figure 4    ADC output SNR (left) and SDR (right) as function of ADC input SNR

## VI. EQUIVALENT MODEL FOR COMPLEX QUANTIZER WITH NOISY INPUT

We now derive the quantizer model in the complex domain. According to (0.37) we may represent output of complex ADC pair as:

$$\tilde{s}_O(m) = \tilde{F}\left(\sigma_N, \tilde{s}_I(m)\right) + \tilde{n}_O(m) = g_O(\sigma_N, \sigma_S) \cdot \tilde{s}_I(m) + \tilde{w}_O(m) + \tilde{n}_O(m) \tag{0.48}$$

where $\tilde{F}(\sigma_N, \tilde{s}_I)$ is the equivalent complex transfer function, $\tilde{w}_O(m)$ is the complex NLD of ADC pair $m$ and $\tilde{n}_O(m)$ is the complex equivalent additive noise of ADC pair $m$:

$$\tilde{F}(\sigma_N, \tilde{s}_I(m)) = F(\sigma_N, \text{Re}(\tilde{s}_I)) + j \cdot F(\sigma_N, \text{Im}(\tilde{s}_I)) \tag{0.49}$$

$$\tilde{w}_O(m) = w_O(2 \cdot m - 1) + j \cdot w_O(2 \cdot m) \tag{0.50}$$

$$\tilde{n}_O(m) = n_O(2 \cdot m - 1) + j \cdot n_O(2 \cdot m) \tag{0.51}$$

From (0.25),(0.26),(0.27),(0.32),(0.33) and (0.36) it follows that the complex quantizer equivalent additive noise and NLD have the following properties:

$$E[\tilde{n}_O(m)] = 0 \quad \text{and} \quad E[\tilde{w}_O(m)] = 0 \quad \text{for any } m \tag{0.52}$$

$$E[\tilde{n}_O(m)^* \cdot \tilde{s}_I(n)] = 0 \quad \text{and} \quad E[\tilde{w}_O(n)^* \cdot \tilde{s}_I(m)] = 0 \quad \text{for any } m \text{ and } n \tag{0.53}$$

$$E[\tilde{n}_O(n)^* \cdot \tilde{w}_O(m)] = 0 \quad \text{for any } m \text{ and } n \tag{0.54}$$

$$E[\tilde{n}_O(m)^* \cdot \tilde{n}_O(n)] = 0 \quad \text{for any } (n \neq m) \tag{0.55}$$

However we cannot assume the NLD is white, $E[\tilde{w}_O(m)^* \cdot \tilde{w}_O(n)] \neq 0$.

According to (0.50) and (0.51):

$$E[|\tilde{n}_O(m)|^2] = E[n_O(2 \cdot m - 1)^2] + E[n_O(2 \cdot m)^2] = 2 \cdot \sigma_{NO}(\sigma_N, \sigma_S)^2 \tag{0.56}$$

$$E[|\tilde{w}_O(m)|^2] = E[w_O(2 \cdot m - 1)^2] + E[w_O(2 \cdot m)^2] = 2 \cdot \sigma_{WO}(\sigma_N, \sigma_S)^2 \tag{0.57}$$

### VII. PERFORMANCE OF MIMO RECEIVER WITH ARRAY OF LOW RESOLUTION ADCS

The receiver that follows finite resolution ADCs has to normalize MRC result by equivalent ADC gain $g_O$. Therefore according to (0.13) the estimate of user signal $x_k$ is equal to,

$$\hat{\tilde{x}}_k = \frac{1}{g_O(\sigma_N, \sigma_S)} \cdot MRC_k(\tilde{S}_O) = \frac{1}{g_O(\sigma_N, \sigma_S)} \cdot \frac{\tilde{g}_k^*}{M \cdot |\tilde{g}_k|^2} \cdot \sum_{m=1}^{M} \tilde{c}_k(m)^* \cdot \tilde{s}_O(m) \tag{0.58}$$

According to the equivalent model (0.37) its output is equal to,

$$\hat{x}_k = x_k + \tilde{n}_{O,k} + \tilde{w}_{O,k} \tag{0.59}$$

where $\tilde{n}_{O,k}$ and $\tilde{w}_{O,k}$ are post-MRC additive white noise and NLD of the *k*-th user respectively.

$$\tilde{n}_{O,k} = \frac{1}{g_O(\sigma_N,\sigma_S)} \cdot \frac{\tilde{g}_k^*}{M \cdot |\tilde{g}_k|^2} \cdot \sum_{m=1}^{M} \tilde{c}_k(m)^* \cdot \tilde{n}_O(m) \quad (0.60)$$

$$\tilde{w}_{O,k} = \frac{1}{g_O(\sigma_N,\sigma_S)} \cdot \frac{\tilde{g}_k^*}{M \cdot |\tilde{g}_k|^2} \cdot \sum_{m=1}^{M} \tilde{c}_k(m)^* \cdot \tilde{w}_O(m) \quad (0.61)$$

Since the equivalent output noise is a white process (see (0.55)):

$$\sigma_{NO,k}^2 = E\left[|\tilde{n}_{O,k}|^2\right] = 2 \cdot \sigma_{NO}(\sigma_N,\sigma_S)^2 / \left(M \cdot g_O(\sigma_N,\sigma_S)^2 \cdot |\tilde{g}_k|^2\right) \quad (0.62)$$

However the NLD is not necessarily white process and in the worst case may sum coherently. The post MRC NLD is minimized when the NLD of each observation $\tilde{w}_O(m)$ is uncorrelated. It is equal to,

$$\min\left(\sigma_{WO,k}^2 = E\left[|\tilde{w}_{O,k}|^2\right]\right) = 2 \cdot \sigma_{WO}(\sigma_N,\sigma_S)^2 / \left(M \cdot \hat{g}_O(\sigma_N,\sigma_S)^2 \cdot |\tilde{g}_k|^2\right) \quad (0.63)$$

The post MRC NLD is maximal when NLD of each observation satisfies:

$$\tilde{w}_O(m) = \left(\tilde{c}_k(m)/\tilde{c}_k(1)\right) \cdot \tilde{w}_O(1) \quad (0.64)$$

When this happens, according to (0.61) and (0.64) the post MRC output NLD is equal to,

$$\tilde{w}_{O,k} = \frac{1}{g_O(\sigma_N,\sigma_S)} \cdot \frac{\tilde{g}_k^*}{M \cdot |\tilde{g}_k|^2} \cdot \sum_{m=1}^{M} \tilde{str}_k(m)^* \cdot \left(\frac{\tilde{c}_k(m)}{\tilde{c}_k(1)}\right) \cdot \tilde{w}_O(1) = \frac{\left(\tilde{g}_k^*/\tilde{c}_k(1)\right) \cdot \tilde{w}_O(1)}{g_O(\sigma_N,\sigma_S) \cdot |\tilde{g}_k|^2} \quad (0.65)$$

Therefore the post MRC NLD variance for the worst case (0.64) is equal to,

$$\max\left(\sigma_{WO,k}^2 = E\left[|\tilde{w}_{O,k}|^2\right]\right) = 2 \cdot \sigma_{WO}(\sigma_N,\sigma_S)^2 / \left(\hat{g}_O(\sigma_N,\sigma_S)^2 \cdot |\tilde{g}_k|^2\right) \quad (0.66)$$

One example of such a worst case is when we deal with a single user (no interference, $K=1$) and the phase shift of each observation *m* satisfies,

$$\tilde{c}_1(m) = \tilde{a}_1(m) \cdot \exp(j \cdot \varphi) \quad \text{where} \quad \tilde{a}_1(m) = (\pm 1 \text{ or } \pm j) \text{ and } \varphi \text{ is any random phase} \quad (0.67)$$

According to (0.6) and (0.5) for single user:

$$\tilde{s}_I(m) = \sum_{k=1}^{K} \tilde{g}_k \cdot \tilde{c}_k(m) \cdot \tilde{x}_k = \tilde{a}_1(m) \cdot \tilde{y} \quad \text{where} \quad \tilde{y} = \tilde{g}_1 \cdot \exp(j \cdot \varphi) \cdot \tilde{x}_1 \quad (0.68)$$

From (0.1) and (0.49) it follows that:

$$\tilde{F}(\sigma_N, -\tilde{x}) = -\tilde{F}(\sigma_N, \tilde{x}) \text{ and } \tilde{F}(\sigma_N, j \cdot \tilde{x}) = j \cdot \tilde{F}(\sigma_N, \tilde{x}) \tag{0.69}$$

Therefore according to (0.31) and (0.50):

$$\begin{aligned}\tilde{w}_O(m) &= F(\sigma_N, \tilde{a}_1(m) \cdot \tilde{y}) - g_O(\sigma_N, \sigma_S) \cdot \tilde{a}_1(m) \cdot \tilde{y} = \\ &= \tilde{a}_1(m) \cdot (F(\sigma_N, y) - g_O(\sigma_N, \sigma_S) \cdot \tilde{y}) = (\tilde{a}_1(m)/\tilde{a}_1(1)) \cdot \tilde{w}_O(1) = (\tilde{c}_k(m)/\tilde{c}_k(1)) \cdot \tilde{w}_O(1)\end{aligned} \tag{0.70}$$

We can see that the expression (0.70) satisfies the worst case NLD definition (0.64).

An example when multi antennas MIMO coefficients $\tilde{c}_1(m)$ satisfy (0.67) according to (0.5) is a uniform linear array with angle of arrival equal to 0, or $\pm \pi/2$, or $\pm \pi/3$.

According to (0.59), (0.62) and (0.66) we may express worst case post MRC output SINAD of MIMO array with finite resolution ADC as:

$$SINAD_{User,Out}(k) = \frac{2 \cdot \sigma_X^2}{\sigma_{NO,k}^2 + \max(\sigma_{WO,k}^2)} = \frac{g_O(\sigma_N, \sigma_S)^2 \cdot |\tilde{g}_k| \cdot \sigma_X^2}{(\sigma_{NO}(\sigma_N, \sigma_S)^2/M) + \sigma_{WO}(\sigma_N, \sigma_S)^2} \tag{0.71}$$

The ADC array Noise Figure (NF) as defined above (0.12) for a single user is now:

$$NF_{User}(k) = \frac{SNR_{User,In}(k)}{SINAD_{User,Out}(k)} = \frac{\sigma_{NO}(\sigma_N, \sigma_S)^2 + M \cdot \sigma_{WO}(\sigma_N, \sigma_S)^2}{g_O(\sigma_N, \sigma_S)^2 \cdot \sigma_N^2} = NF_{MAX}(\sigma_N, \sigma_S) \tag{0.72}$$

As we can see from this expression the worst case NF is independent of the user index $k$ and always equal to constant value $NF_{MAX}(\sigma_N, \sigma_S)$.

According to (0.11) $SNR_{ADC,In} = SNR_{\Sigma,In}/M$. Therefore according to (0.10) we may also express worst case NF as function of ADC input SF and cumulative input SNR.

$$NF_{MAX}(SF, SNR_{\Sigma,In}) = NF_{MAX}\left(\sigma_N = R\bigg/\sqrt{SF \cdot (1 + (SNR_{\Sigma,In}/M))}, \sigma_S = \sigma_N \cdot \sqrt{SNR_{\Sigma,In}/M}\right) \tag{0.73}$$

Let us define the worst case cumulative output SINAD as sum of worst case post MRC output SINAD of all users:

$$SINAD_{\Sigma,In} = \sum_{k=1}^{K} SINAD_{User,Out}(k) \tag{0.74}$$

According to (0.72) and (0.74) we may express it as function of $(SF, SNR_{\Sigma,In})$:

$$SINAD_{\Sigma,In}(SF, SNR_{\Sigma,In}) = \sum_{k=1}^{K} \frac{SNR_{User,In}(k)}{NF_{MAX}(SF, SNR_{\Sigma,In})} = \frac{SNR_{\Sigma,In}}{NF_{MAX}(SF, SNR_{\Sigma,In})} \qquad (0.75)$$

From (0.30), (0.20), (0.31) and (0.40) it follows that for 1-bit ADC:

$$\lim_{\sigma_S \to 0}(g_O(\sigma_N, \sigma_S)) = \lim_{\sigma_S \to 0}\left(\int_{x=-\infty}^{+\infty} \frac{F(\sigma_N, \sigma_S \cdot x)}{\sigma_S} \cdot x \cdot p_S(x) \cdot dx\right) = \int_{x=-\infty}^{+\infty} \lim_{\sigma_S \to 0}\left(\frac{F(\sigma_N, \sigma_S \cdot x)}{\sigma_S \cdot x}\right) \cdot x^2 \cdot p_S(x) \cdot dx =$$
$$= \lim_{\sigma_S \to 0}(F(\sigma_N, \sigma_S \cdot x)/(\sigma_S \cdot x)) = \partial F(\sigma_N, x)/\partial x\big|_{x=0} = \partial erf\left(x/\sqrt{2 \cdot \sigma_N^2}\right)/\partial x\big|_{x=0} = \sqrt{2/(\pi \cdot \sigma_N^2)} \qquad (0.76)$$

Analogically from (0.20), (0.31) and (0.40) it follows that for 1-bit ADC:

$$\lim_{\sigma_S \to 0}\left(\sigma_{WO}(\sigma_N, \sigma_S)^2\right) = 0 \qquad \lim_{\sigma_S \to 0}\left(\sigma_{NO}(\sigma_N, \sigma_S)^2\right) = 1 \qquad (0.77)$$

Therefore from (0.72) it follows that:

$$\lim_{SNR_{\Sigma,In} \to 0}\left(NF_{MAX}(G, SNR_{\Sigma,In})\right) = \lim_{\sigma_S \to 0}\left(\frac{\sigma_{NO}(\sigma_N, \sigma_S)^2 + M \cdot \sigma_{WO}(\sigma_N, \sigma_S)^2}{g_O(\sigma_N, \sigma_S)^2 \cdot \sigma_N^2}\right) = \frac{\pi}{2} \approx 1.96 dB \qquad (0.78)$$

which confirms the well-known conclusion [23] and [41] that in the low SNR regime, the performance degradation caused by 1 bit ADC converges to 1.96dB.

We used numerical integration to evaluate the worst case ADC array NF (0.73), and the cumulative output SINAD (0.75). The numerical integration interval was chosen from minus to plus $5 \cdot \sqrt{\sigma_S^2 + \sigma_N^2}$ with step $0.01 \cdot \min(\sigma_S, \sigma_N)$. The SF was optimized to minimize NF:

$$SF_{Optim} = \arg\min_{SF}\left(NF_{MAX}(SF, SNR_{\Sigma,In})\right) \qquad (0.79)$$

The optimal SF was obtained by searching SF from 0 to 30dB with a step size of 0.1dB.

Figure 5 and Figure 6 summarize various aspects of the results of this evaluation for ADC resolution from 1 to 5 bits and for observations number 1,10,100,1000 and 10000.

We may see from these figures that when the cumulative input SNR is low the increase of cumulative input SNR causes linear increase in cumulative output SINAD, and the NF stays at a fixed low level. For 1-bit ADC this level is 1.96 dB, as was predicted in (0.78). However, as the cumulative input SNR grows beyond a certain limit any additional increase in cumulative input SNR starts to decrease the cumulative output SINAD, causing dramatic growth of NF. The reason for this is that when SNR is low, the NLD is mild and when SNR is high the NLD becomes dominant as we had seen in Figure 4. The higher the ADC resolution and the greater the number of observations, the higher the threshold for cumulative input SNR below which a conventional MRC can operate without significant NF degradation.

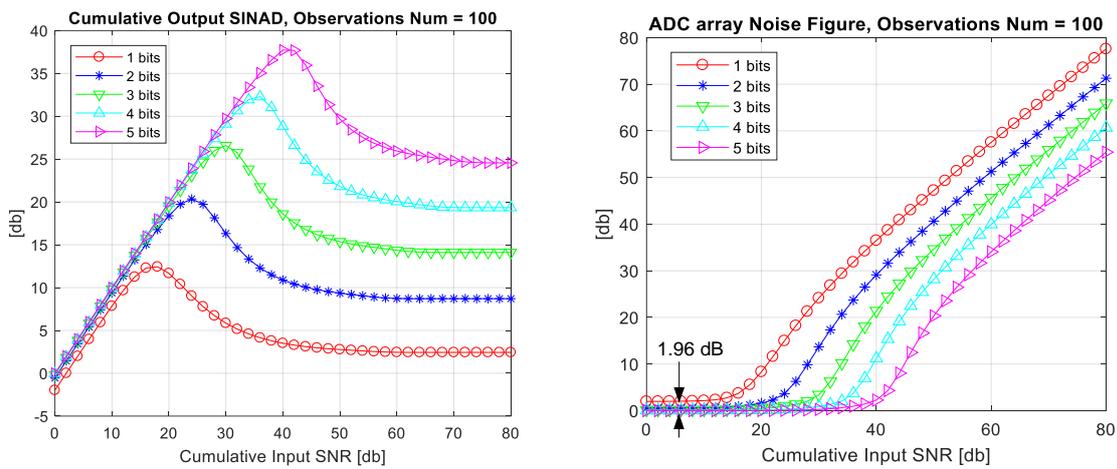

**Figure 5**   The Cumulative output SINAD (0.75) (left) and NF (0.73) (right) as function of Cumulative input SNR for *M* = 100

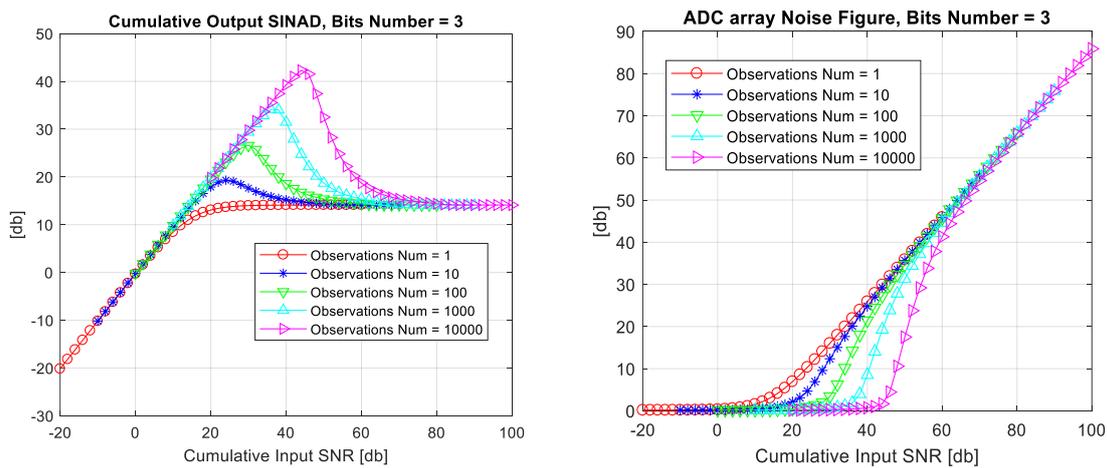

Figure 6   The Cumulative output SINAD (0.75) (left) and NF (0.73) (right) as function of Cumulative input SNR for 3bit ADC

Similar curves of Single User Post MRC output SINAD as function of ADC input SNR were presented in [7]. However the results therein were obtained from numerical simulations, whereas our results are from numerical integration of analytical expressions.

## VIII. ADC RESOLUTION DETERMINATION METHODOLOGY

Our equivalent model provides a practical design rule to determine the ADC resolution. Say we allow a 3 dB performance degradation when an infinite-resolution ADC array is replaced by a finite-resolution one which means maximal NF equals 3dB. We can compute the maximal allowable value of cumulative input SNR accordingly.

$$SNR_{\Sigma,In,3dB} = \underset{SNR_{\Sigma,In}}{\text{Solve}}\left(\underset{SF}{\min}\left(NF_{MAX}\left(SF, SNR_{\Sigma,In}\right)\right) = 3dB\right) \tag{0.80}$$

Where $\underset{x}{\text{Solve}}\left(f(x) = A\right)$ means finding the value of x that fulfills equality $f(x) = A$ and NF given by equation (0.73)

We apply numerical integration to evaluate the above cumulative input SNR 3dB threshold according to (0.80) for Gaussian input signal distribution, for different ADC resolutions and different number of observations. The results are shown in Figure 7. As we may see from this figure, for a given number of observations, a desired cumulative SNR requires a certain number of ADC bits.

Our methodology uses tables of the type illustrated by Figure 7. First we set the number of observations (the number of antennas) and the maximal possible cumulative SNR. Second we use Figure 7 to find the curve that laying just above the intersection point of the maximal possible cumulative SNR on the *x* axes and the observations number on the *y* axes. The number of bits associated with the curve is required ADC resolution.

As an example our methodology, assuming we have 10,000 receive antenna elements and maximal possible cumulative SNR is equal to 40dB, then according to Figure 7, a 3-bit ADC is sufficient.

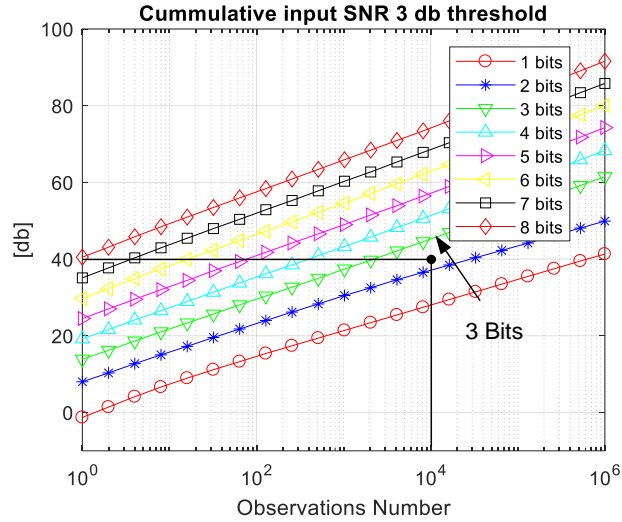

Figure 7    The Cumulative input SNR threshold

## IX. NUMERICAL SIMULATION RESULTS

To confirm the theoretical analysis, we simulate a MIMO uplink receiver system with a single user and a base station equipped with 100 antennas. The modulation is OFDM with QAM64 on each subcarrier. MRC combining is employed. We simulate ADC with 1,2,3,4 and 5 bits of resolution. The ADC input SF was optimized according to(0.79). The channel is LOS defined by (0.5).

Figure 8 presents simulation results for a worst case channel when the angle of arrival is 0° and for an average channel when the angle of arrival is a random variable uniformly distributed from 0 to $\pi$.

We see that when the cumulative input SNR (post MRC input SNR) is low and nonlinear distortion is mild, increasing the input SNR improves the performance. However, after the cumulative input SNR exceeds a certain threshold non linear distortion becames dominant and further increase of cumulative input SNR causes performance degradation. The higher the ADC resolution, the greater the value of this threshold. For the worst case scenario, the value of this threshold (at least for the 1-bit, 2-bit, 3-bit cases) matches our prediction in Figure 7. The performance on average channels is somewhat better, but the same threshold applies.

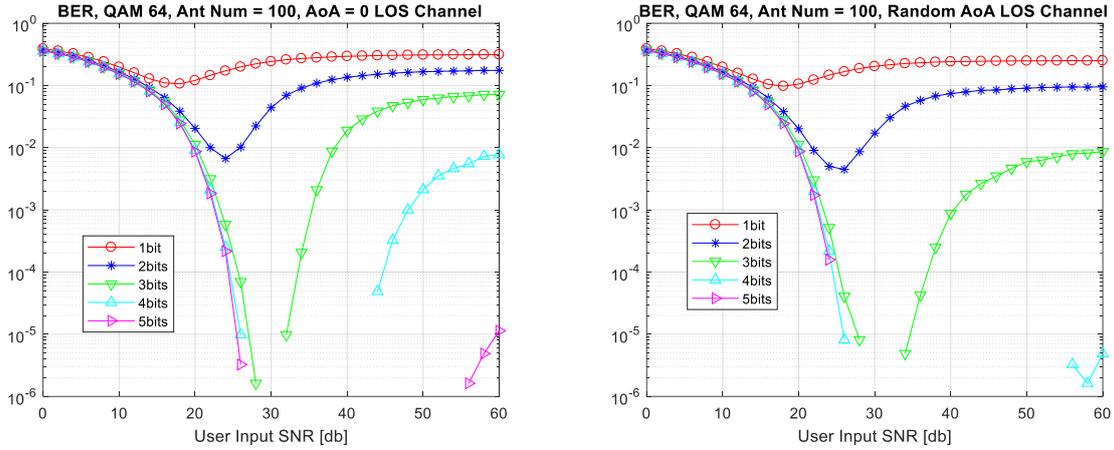

Figure 8    The BER for worst case (left) and average channel (right).

## X.  FUTURE WORK

In this paper we focused on the LOS (flat fading) channel to simplify the analysis and illustrate key concepts. Future work aims to extend the analysis to frequency-selective multipath channels. We speculate that a real-valued multi-path channel has the following effects on the NF of an array of low resolution ADCs:

In Section VII we show that ADC array NF reaches its maximum when there is no interference. Multipath introduces inter-symbol interference that may serve as a dither which improves linearity of ADC transfer function and improves ADCs array NF.

- Multipath increases time domain spreading of received signal energy which may have a positive effect on NF.

- For multi antenna MIMO, multipath may cause non-uniform distribution of desired signal energy between antennas which reduces NF. This may happen only if the system bandwidth is not large enough to absorb all energy fluctuation. To take this problem into account we propose to define the worst case effective antenna number as:

$$\hat{M} = \sum_{m=1}^{M} \sigma_S(m)^2 \Big/ \max\left(\sigma_S(m)^2\right) \tag{0.81}$$

where $\sigma_S(m)^2$ is the desired signal variance at antenna $m$.

We could use the effective antenna number instead of actual antenna number to determine ADC resolution. We define the channel efficency coeficient as: $\alpha = \hat{M}/M$. Then effective antenna number is equal to $\hat{M} = \alpha \cdot M$. The channel efficency coeficient $\alpha$ is determined by channel measurements and should be part of the channel model.

Furthermore, for multi-user uplink transmission where the base station is equipped with an antenna array, power control by terminals is necessary to attain the optimal SNRs at ADC inputs calculated in this paper. Such schemes are for further study.

## XI. CONCLUSIONS

We presented a novel equivalent model of quantizer with noisy inputs, which can be applied to obtain a design rule for setting the ADC resolution in a MIMO system equipped with an array of low resolution ADCs.

- Our analysis shows the critical SNR (tabulated in, e.g. Figure 7) below which we may use a conventional receiver that does not take NLD into account without suffering significant performance degradation. This result was confirmed by our simulations.

- Above this critical SNR, there are two methods for dealing with the resulting NLD:

    ▪ We can lower the input SNR by dithering, as proposed by [50]. However, our model gives the relationship between the equivalent transfer function and the noise PDF, leading to an improved dither with an optimal PDF. In a companion patent [54] and paper [56] we present how our model may be used to design optimal dither for the ADC in the receiver, and also for the DAC in the transmitter.

    ▪ We can develop new receivers that take into account the NLD caused by an ADC with insufficient resolution. In other words, NLD suppressing equalization, as proposed by [32]-[33]. However, in our model the NLD is expressed as a function of the desired input signal only, leading to efficient low-complexity NLD compensation schemes. In an accompanying paper [55] we present an example of such receivers.